\documentclass[aps,PRApplied,twocolumn,amsmath,superscriptaddress, showpacs,floatfix,reprint]{revtex4-2}
\usepackage{amsmath, nccmath}% need for subequations
\usepackage{amssymb}
\usepackage{bm}
\usepackage{braket}
\usepackage{natbib}
\usepackage{leftidx}
\usepackage{graphicx}    % need for figures
\usepackage{verbatim}   %useful for program listings
\usepackage{color}      %use if color is used in text
\usepackage{subfigure}  %use for side-by-side figures
\usepackage{hyperref}   %use for hypertext links, including those to external documents and URLs
\usepackage{dcolumn}    %use for double column
\usepackage{textcomp}
\usepackage{float}
\usepackage{threeparttable}
\usepackage{titlecaps}% http://ctan.org/pkg/titlecaps
\usepackage{multirow}
\usepackage{lipsum}% http://ctan.org/pkg/lipsum
\hypersetup{
	colorlinks=true,
	citecolor=blue,
	filecolor=black,
	linkcolor=blue,
	urlcolor=cyan
}
\usepackage{mathastext}
\usepackage{mathptmx}
\usepackage{enumitem}
\DeclareGraphicsExtensions{.png,.pdf,.tif}
\usepackage{pict2e}
\usepackage[dvipsnames]{xcolor}
\usepackage[normalem]{ulem}
\begin{document}

\title{Magnetocrystalline Anisotropy and Magnetocaloric Effect Studies on the Room-temperature 2D Ferromagnetic Cr$_4$Te$_5$}
\author{Shubham Purwar}
\author{Susmita Changdar}
\author{Susanta Ghosh}%
\author{S.\ Thirupathaiah}
\email{setti@bose.res.in}
\affiliation{Department of Condensed Matter and Materials Physics, S. N. Bose National Centre for Basic Sciences, Kolkata, West Bengal, India, 700106.}

%\\This line break forced% with \\
%

\date{\today}% It is always \today, today,

\begin{abstract}

 We present a thorough study on the magnetoanisotropic properties and magnetocaloric effect in the layered ferromagnetic Cr$_4$Te$_5$ single crystals by performing the critical behaviour analysis of magnetization isotherms. The critical exponents $\beta$=0.485(3), $\gamma$=1.202(5), and $\delta$=3.52(3) with a Curie temperature of $T_C \approx 340.73(4)$ K are determined by the modified Arrott plots. We observe a large magnetocrystalline anisotropy K$_u$=330 kJ/$m^3$ at 3 K which gradually decreases with increasing temperature. Maximum entropy change -$\Delta S_{M}^{max}$ and the relative cooling power (RCP) are found to be 2.77 $J/kg-K$ and 88.29 $J/kg$, respectively near $T_C$ when the magnetic field applied parallel to $\it{ab}$-plane. Rescaled -$\Delta S_M (T, H)$ data measured at various temperatures and fields collapse into a single universal curve,  confirming the second order magnetic transition in this system. Following the renormalization group theory analysis, we find that the spin-coupling is of 3D Heisenberg-type, $\{d:n\}=\{3:3\}$,  with long-range exchange interactions decaying as $J (r) = r^{-(d+\sigma)}= r^{-4.71}$.
 \end{abstract}

\maketitle
\section{\label{sec:level1}Introduction}

Investigation of the low dimensional magnetic materials with ferromagnetic ordering at room temperature has gained a lot of research interests in recent days due to their potential applications in the magnetic refrigeration and spintronic devices~\cite{deng2018gate,wang2019mnx,geim2013van,PhysRevLett.107.217202} for which the materials with strong magentocrystalline anisotropy are crucial~\cite{liu2022interlayer,soumyanarayanan2016emergent}. In this regard, the van der Waals (vdW) magnets are of great interest from the fundamental science and advanced technological point of view due to their peculiar 2-dimensional (2D) magnetic properties~\cite{gong2017discovery,seyler2018ligand,fei2018two,gibertini2019magnetic} and strong magnetocrystalline anisotropy~\cite{PhysRevB.93.134407}. Usually, the Heisenberg-type ferromagnet does not exist with intrinsic long-range magnetic ordering at finite temperature in the 2D limit due to dominant thermal fluctuations~\cite{PhysRevLett.17.1133}. But the single domain anisotropy or the exchange anisotropy can remove these limitations and allow the long-range magnetic ordering to obtain a 2D Ising-like ferromagnet~\cite{Huang2017}. In this way, 2D ferromagnetism has been found experimentally in many vdW materials such as Cr$_2$Ge$_2$Te$_6$ (T$_C\approx$ 61 K)~\cite{PhysRevB.96.054406}, Cr$_2$Si$_2$Te$_6$ (T$_C\approx$ 32 K)~\cite{PhysRevB.92.144404}, Fe$_3$GeTe$_2$ (T$_C\approx$ 215 K)~\cite{fei2018two}, and CrI$_3$ (T$_C \approx$ 45 K)~\cite{Huang2017},  but the long-range FM ordering far below the room temperature limiting their usage in technology. However, a very few systems such as MnP (T$_C\approx$ 303 K) show room temperature 2D ferromagnetism in the bulk phase~\cite{Sun2020}. Thus, searching for new 2D room-temperature FM materials with large magnetocrystalline anisotropy is crucial for the realization of potential technological applications~\cite{wang2019mnx,deng2018gate}. Further, magnetocaloric effect (MCE) in room-temperature FM vdW materials with maximum entropy change -$\Delta S_{M}^{max}$  is of recent research interests due to their potential applications in the environmental friendly magnetic refrigeration~\cite{magnetochemistry7050060}.  There exists quite a few 2D FM systems showing significant -$\Delta S_{M}^{max}$ such as Fe$_{3-x}$GeTe$_2$ (1.1 $J/kg-K$ at 5 T)~\cite{doi:10.1021/acs.inorgchem.5b01260},  Cr$_5$Te$_8$ (1.6 $J/kg-K$ at 5 T)~\cite{PhysRevB.100.245114}, Cr$_2$Si$_2$Te$_6$ (5.05 $J/kg-K$ at 5 T), and  Cr$_2$Ge$_2$Te$_6$ (2.64 $J/kg-K$ at 5 T)~\cite{PhysRevMaterials.3.014001}.

%CrI$_3$ shows -$\Delta S_{M}^{Max}$ of 4.24 and 2.68 J-Kg$^{-1}$ K$^{-1}$ for H$\parallel c$ and H$\parallel ab$, respectively~\cite{PhysRevB.97.174418}. The magnetocrystalline anisotropy constant K$_u$ is larger in Cr$_2$Si$_2$Te$_6$ compared to Cr$_2$Ge$_2$Te$_6$, resulting in an increased rotational magnetic entropy change at T$_C$\cite{PhysRevMaterials.3.014001}.

  \begin{figure}[t]
    \centering
    \includegraphics[width=\linewidth]{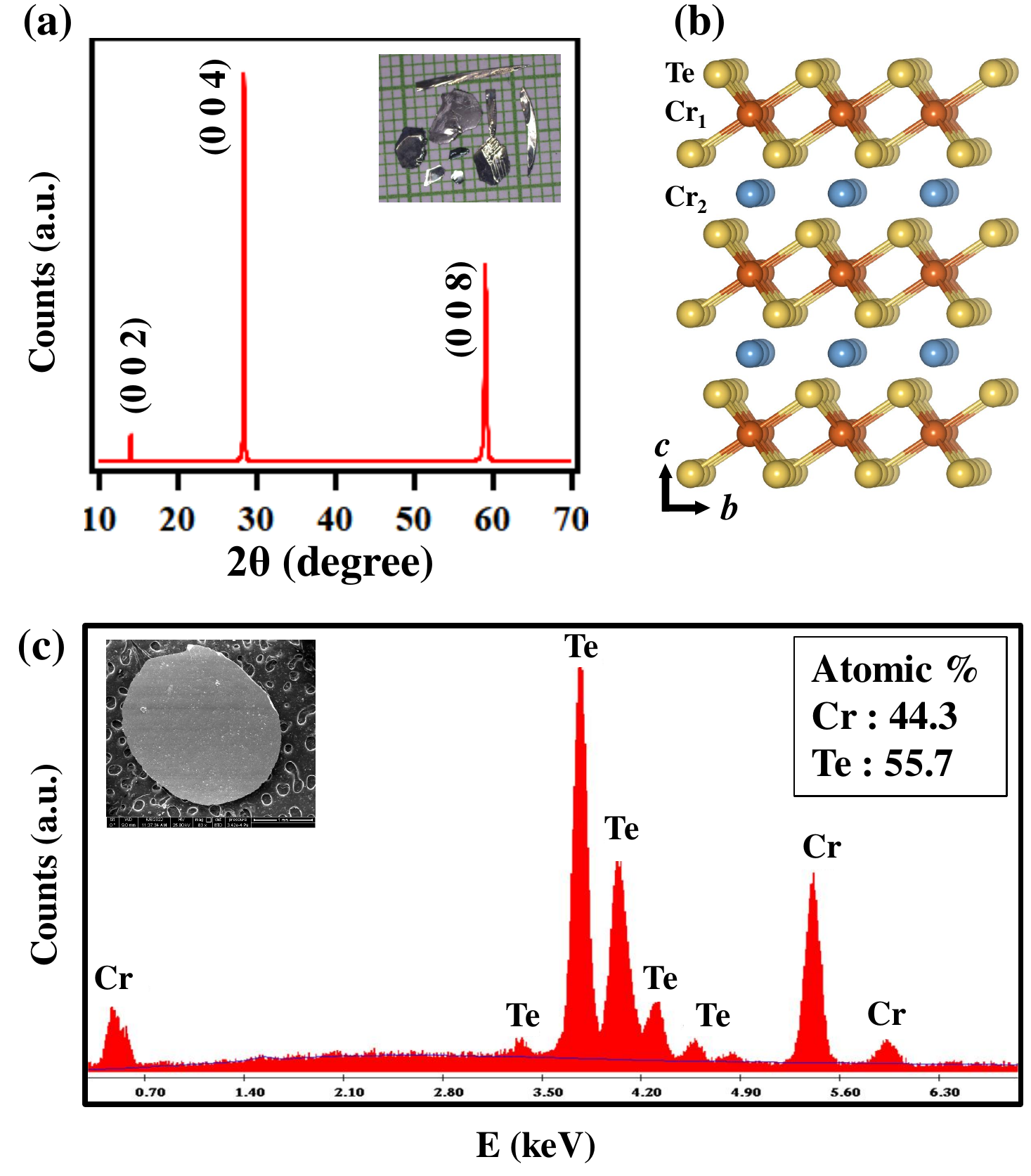}
    \caption{(a) X-ray diffraction pattern of Cr$_4$Te$_5$ single crystals. Inset in (a) shows photographic image of single crystals. (b). Schematic diagram of Cr$_4$Te$_5$ crystal structure. (c) Energy dispersive X-ray spectroscopy and scanning electron microscopy image (inset) of Cr$_4$Te$_5$ single crystal showing the chemical composition and crystal morphology, respectively.}
    \label{1}
\end{figure}

A Recent theoretical study suggested that the layered Cr$_x$Te$_y$ systems are potential candidates to realize the much-anticipated room-temperature 2D ferromagnetism in the bulk~\cite{PhysRevMaterials.2.081001}. Since then, variety of Cr$_x$Te$_y$ compounds have been grown experimentally and studied for their peculiar 2D ferromagnetism, including,  CrTe~\cite{ETO200116}, Cr$_2$Te$_3$~\cite{wang2018ferromagnetic}, Cr$_3$Te$_4$~\cite{hessen1993hexakis}, Cr$_4$Te$_5$~\cite{Zhang2020},  and Cr$_5$Te$_8$~\cite{PhysRevB.100.024434}. Out of these, Cr$_4$Te$_5$ has been reported to show highest T$_C$ $\approx$ 318 K among this family of compounds, to date in the bulk form. Generally, Cr$_x$Te$_y$ compounds possess alternating stacks of CrTe$_2$ layers intercalated by the Cr layers (excess) along the $c$-axis~\cite{Dijkstra_1989}. Importantly, it is that the intercalated Cr concentration plays a crucial role in the crystal structure formation, magnetic structure, and transport properties of these compounds\cite{Huang2004, Huang2006, Wontcheu2008, IPSER1983265}.

In this contribution, we present magnetic properties, critical behaviour analysis, and anisotropic magnetocaloric effect associated with room temperature ferromagnetism in Cr$_4$Te$_5$ single crystals. Magnetization $M(T)$ data suggest ferromagnetic ordering at a T$_C$ of $\approx$ 338 K with strong magnetic anisotropy below T$_C$ between the field applied parallel to $\it{ab}$-plane and parallel to $c$-axis. As a result, we find a large temperature dependent magnetocrystalline anisotropy (K$_u$=330 kJ/m$^3$) at 3 K with $\it{ab}$-plane as the easy magnetization plane.  Further, the universal scaling of magnetocaloric effect suggests  a second order magnetic phase transition in this system.  The critical exponents derived from the magnetocaloric effect are in good agreement with the values obtained from the critical analysis. The renormalization group theory analysis suggests 3D Heisenberg-type magnetic interactions in this system.

\section{Experimental details}
High quality single crystals of Cr$_4$Te$_5$ were grown by the chemical vapor transport (CVT) technique with iodine as a transport agent. Stoichiometric mixtures of Cr (99.99\%, Alfa Aesar) and Te  (99.99\%, Alfa Aesar) powders were mixed in a molar ratio of 5:5, added with iodine (3 mg/cc),  and sealed into an evacuated quartz tube under Ar atmosphere. The tube was placed in a gradient-temperature horizontal two-zone tube furnace for about 15 days with one end, containing powder mixture (source), maintained at 1000$^o$ C and the other end kept at 820$^o$ C (sink), as per the procedure described earlier~\cite{hashimoto1971magnetic}. The as-grown single crystals were large in size ($5\times5$ mm$^2$) and were looking  shiny. Photographic image of a typical single crystal is shown in the inset of Fig.~\ref{1}(a). X-ray diffraction (XRD) was performed on the single crystals using Rigaku X-ray diffractometer (SmartLab, 9kW) with Cu K$_\alpha$ radiation of wavelength 1.54059 \AA. Surface morphology and elemental analysis of the single crystals were done using the scanning electron microscopy (SEM) and energy dispersive X-ray spectroscopy (EDXS), respectively. The EDXS measurements suggest an actual chemical composition of Cr$_{3.98}$Te$_5$. Hereafter, for convenience, we use nominal composition formula of Cr$_4$Te$_5$ wherever required. Magnetic measurements [$M(T)$ and $M(H)$] were carried out using physical property measurement system (9 Tesla-PPMS, DynaCool, Quantum Design).

\section{Results and Discussion}

Figure~\ref{1}(a) shows X-ray diffraction (XRD) pattern of Cr$_4$Te$_5$ single crystal. In the XRD pattern, we find intensity peaks of $(00l)$ Bragg plane, indicating the crystal growth plane along the $c$-axis. We further notice that the $(00l)$ Bragg peak position is shifted to lower 2$\theta$ values compared to the other Cr$_x$Te$_y$ compounds~\cite{PhysRevB.96.134410,ZHANG2018798} as excess Cr atoms are intercalated between two CrTe$_2$ layers, leading to increase in the lattice parameters~\cite{coughlin2021van}.  Thus, the intercalated Cr atoms (Cr-vacancies) sandwiched between two CrTe$_2$ layers create van der Waals gap as schematically shown in Fig.~\ref{1}(b).

\begin{figure}[t]
    \centering
    \includegraphics[width=\linewidth]{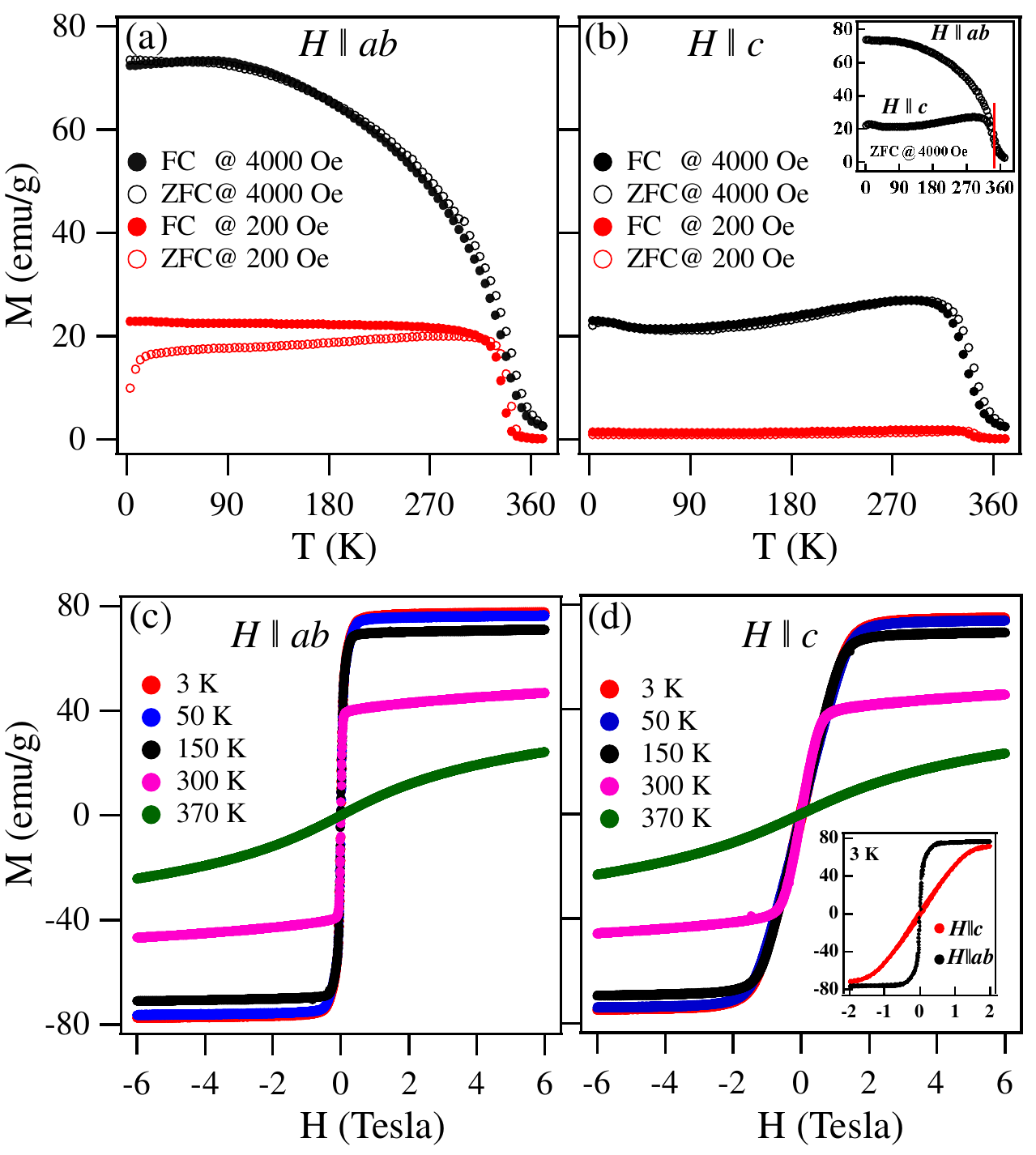}
    \caption{Temperature dependent magnetization $M(T)$ data measured in the zero-field-cooled and field-cooled modes with the magnetic fields (H = 200 and 4000 Oe) applied in the $\it{ab}$-plane  (a) and along the $c$-axis (b).  Field dependent magnetization $M(H)$ data measured for  $H\parallel ab$ (c) and $H\parallel c$ (d) at various sample temperatures.}
    \label{2}
\end{figure}

\subsection{Magnetization Measurements}

To explore the magnetic properties of Cr$_4$Te$_5$, magnetization as a function of temperature $M(T)$ and field $M(H)$ were measured with the field applied parallel to $\it{ab}$-plane ($H\parallel ab$) and $c$-axis ($H\parallel c$). Figs.~\ref{2}(a) and ~\ref{2}(b) show $M(T)$ of Cr$_4$Te$_5$ measured with magnetic fields of 200 and 4000 Oe for both $H\parallel ab$ and $H\parallel c$ orientations, respectively. Paramagnetic (PM) to ferromagnetic (FM) transition is clearly visible at a Curie temperature of $T_C \approx$ 338 K when measured with 200 Oe for both the orientations in zero-field-cooled (ZFC) and field-cooled (FC) modes.  The Curie temperature 338 K found in this study is higher than any known composition in the Cr$_x$Te$_y$ family in their bulk phase~\cite{Zhang2020,ZHANG2022168770}.   Splitting in the $M(T)$ curves between ZFC and FC data at $T_C$ is attributed to the competition between FM and AFM phases at low temperatures~\cite{zhang2020tunable, HUANG20081099}. Interestingly, as shown in the inset of the Fig.~\ref{2}(b), below $T_C$, the  magnetization $M(T)$ increases with decreasing temperature for $H\parallel ab$, while it decreases with decreasing temperature for $H\parallel c$. This observation clearly hints at a significant magnetic anisotropy in this system which is temperature dependent~\cite{PhysRevB.100.024434}.

% and the low temperature magnetization at 3 K is 3 times higher for $H\parallel ab$ (73.5 emu/g) compared to $H\parallel c$ (22 emu/g).

Next, Figs.~\ref{2}(c) and \ref{2}(d) show the magnetization isotherms $M(H)$ measured for $H\parallel ab$ and $H\parallel c$ at various temperatures, respectively.  In contrast to the other Cr$_x$Te$_y$ systems where the easy-axis is along $c$-axis~\cite{zhang2020tunable,Yan_2019,MONDAL201927}, Cr$_4$Te$_5$ has the easy-axis on the $\it{ab}$-plane as the magnetization saturation occurs at an applied field of 0.5 T for $H\parallel ab$,  while 1.9 T is required to attain the magnetization saturation for $H\parallel c$ at 3 K. This observation is in agreement with a previous study on Cr$_4$Te$_5$~\cite{Zhang2020}.  On the other hand, as can be seen from Figs.~\ref{2}(c) and \ref{2}(d), the saturation magnetization ($M_{s}$) gradually decreases with increasing temperature and at $\approx$ 338 K the long-range magnetic order is suppressed in both orientations. Inset in Fig.~\ref{2}(d) demonstrates the magnetization isotherm comparison between $H\parallel ab$ and $H\parallel c$ orientations at 3 K.

\begin{figure}[t]
    \centering
    \includegraphics[width=\linewidth]{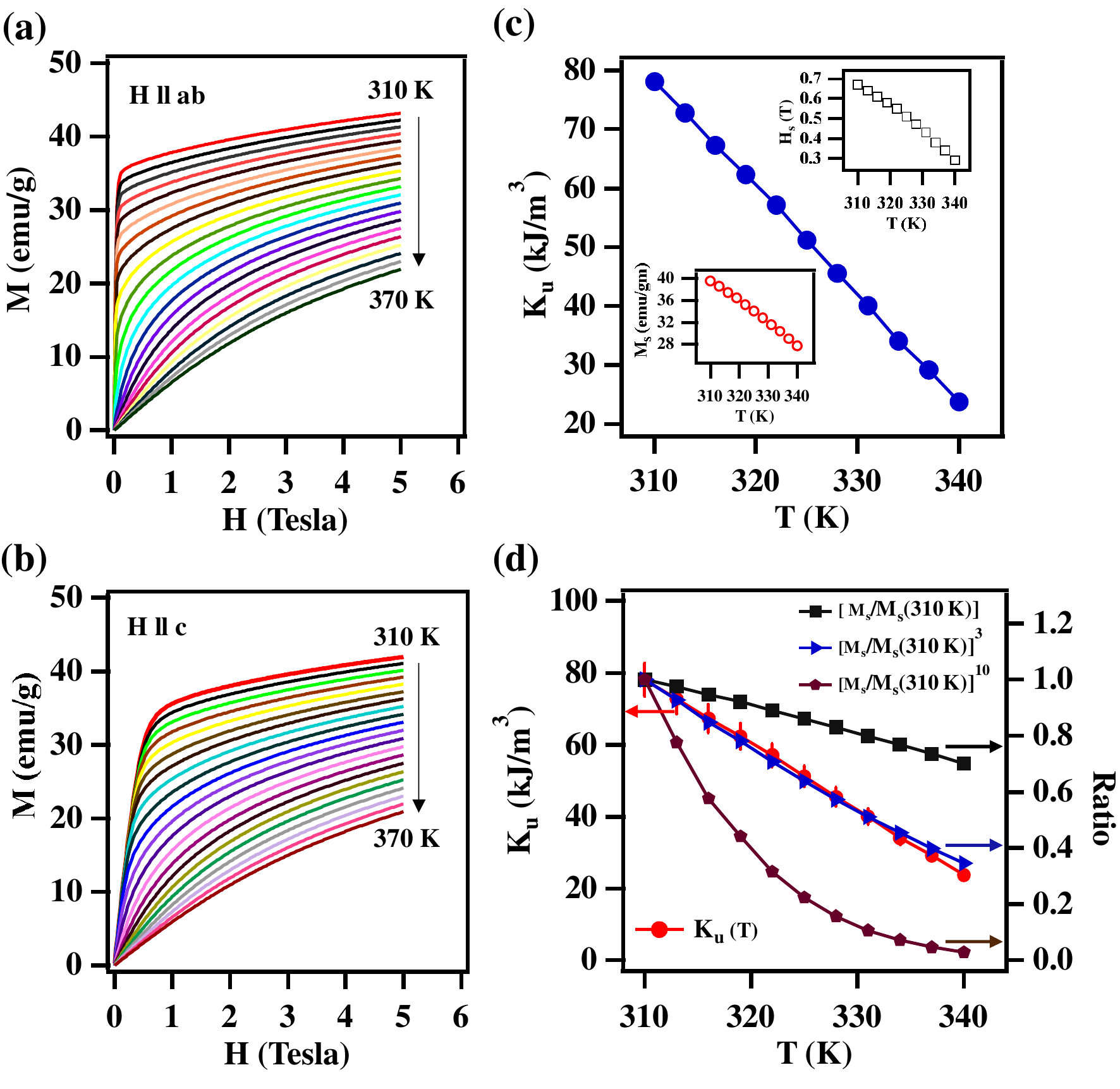}
    \caption{Magnetization isotherms measured around $T_C$ for (a) $H\parallel ab$ and  (b) $H\parallel c$.   (c) Temperature dependent magnetocrystalline anisotropy ($K_u$). Bottom inset in (c) shows the saturation magnetization M$_s$ and top inset in (c) shows saturation magnetic field H$_s$  estimated below $T_C$. (d) Ratios of [M$_s$/M$_s$(310 K)]$^{n(n+1)/2}$ with n =1, 2, and 4 (right-axis) overlapped with magnetocrystalline anisotropy (left-axis).}
    \label{3}
\end{figure}

\subsection{Magnetocrystalline Anisotropy}

We studied the magnetocrystalline anisotropy (K$_u$) using the Stoner-Wolfarth model~\cite{stoner1948mechanism} with the help of magnetization isotherms shown in the Figs.~\ref{3}(a) and \ref{3}(b) for both  $H\parallel ab$ and $H\parallel c$, respectively, measured around T$_C$. The magnetocrystalline anisotropy is calculated by using the formula $2K_u/M_s = \mu_0H_s$, where $\mu_0$ is the vacuum permeability, M$_s$ is the saturation magnetization, and H$_s$ is the field at which saturation magnetization occur~\cite{cullity2011introduction}. Both H$_s$ and M$_s$ are considered experimentally from the isotherms shown in Figs.~\ref{3}(a) and \ref{3}(b) to obtain K$_u$. Fig.~\ref{3}(c) depicts the temperature dependent K$_u$ estimated from the experimental data. We find K$_u$=78.11 kJ/m$^3$ at T = 310 K which gradually decreases with increasing temperature and reaches to 23.73 kJ/m$^3$ at $T_C$ (338 K). On the other hand,  we estimate K$_u$ $\approx$ 330 kJ/m$^3$ at 3 K that is much larger than the K$_u$ values of other 2D magnetic systems such as CrBr$_3$ ($\approx$ 86 kJ/m$^3$ at 5 K)~\cite{PhysRevMaterials.2.024004}, Cr$_2$Ge$_2$Te$_6$ ($\approx$ 20 kJ/m$^3$ at 2 K)~\cite{PhysRevMaterials.3.014001}, and Cr$_2$Si$_2$Te$_6$ ($\approx$ 65 kJ/m$^3$ at 2 K)~\cite{PhysRevMaterials.3.014001} and comparable to the K$_u$ values of CrI$_3$ (300 kJ/m$^3$  at 5 K)~\cite{PhysRevMaterials.2.024004} and Fe$_4$GeTe$_2$ (250 kJ/m$^3$  at 80 K)~\cite{PhysRevMaterials.3.014001}. Though Fe$_3$GeTe$_2$ shows extremely large magnetocrystalline anisotropy of 1460 kJ/m$^3$ at 2 K~\cite{leon2016magnetic}, it's Curie temperature is much below the room temperature (T$_C$=220 K)~\cite{LeonBrito2016}. Thus, our studied sample of Cr$_4$Te$_5$ having large K$_u$ value of $\approx$ 330 kJ/m$^3$ above room temperature looks promising candidate from the technological point of view.

Since the magnetic anisotropy expectation value $<K^n>$ is directly proportional to $M_s^{n(n+1)/2}$, as per the classical theory of magnetism~\cite{PhysRev.96.1335,carr1958temperature}, we plotted $[M_{s}(T)/M_{s}(310)]^{n(n+1)/2}$ for n=1, 2, and 4. Here, n=1 represents intrinsic anisotropy, n=2 represents uniaxial anisotropy, and n=4 represents cubic anisotropy, giving rise the exponents 1, 3, and 10, respectively. Thus, in Fig.~\ref{3}(d), we plotted $[M_{s}(T)/M_{s}(310)]$,  $[M_{s}(T)/M_{s}(310)]^3$, and $[M_{s}(T)/M_{s}(310)]^{10}$ ratios as a function of temperature. Most importantly, in Fig.~\ref{3}(d),  the overlapped K$_u$(T) of Fig.~\ref{3}(c) matches very well with $[M_{s}(T)/M_{s}(310)]^3$ ratio, confirming the dominant uniaxial anisotropy in this system that is strongly temperature dependent. The temperature dependent K$_u$ is found to be originated from the fluctuating local spin clusters those are activated from the thermal energy~\cite{PhysRev.96.1335,carr1958temperature}.  A recent study on Cr$_5$Te$_8$ shows temperature dependent uniaxial anisotropy but mixed with a small component of cubic anisotropy as well~\cite{PhysRevB.100.245114}. Nevertheless, our observations suggest that Cr$_4$Te$_5$ is a stronger magnetocrystalline anisotropic system compared to it's sister compound Cr$_5$Te$_8$.

\begin{figure*}[t]
    \centering
    \includegraphics[width=0.95\linewidth]{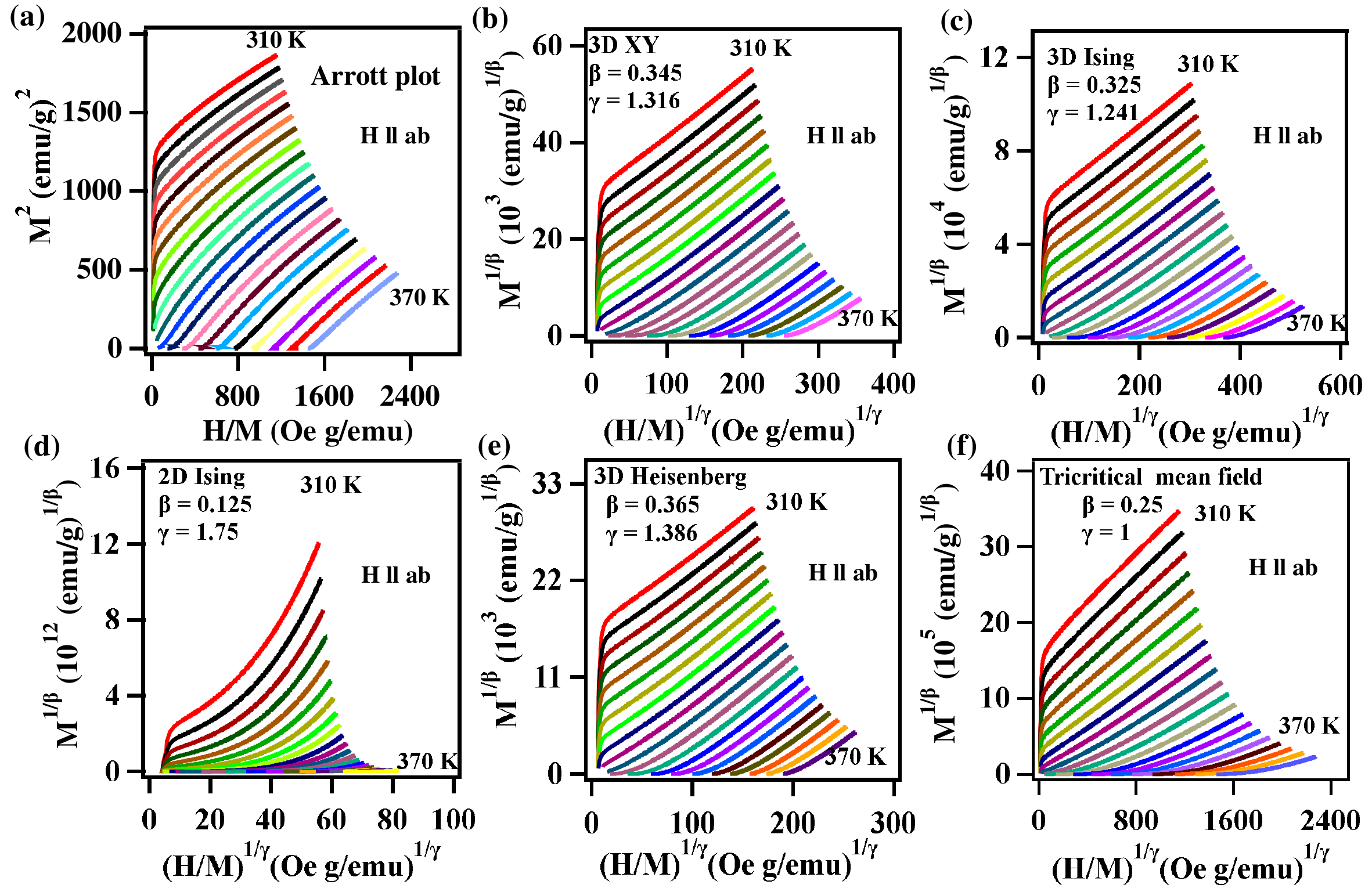}
        \caption{ (a) Arrott plot ($M^2$ vs $H/M$). Modified Arrott plot from (b) 3D XY model, (c) 3D Ising model, (d) 2D Ising model, (e) 3D Heisenberg model,  and (f) Tricritical meanfield model.}
    \label{4}
\end{figure*}

\subsection{Critical Behaviour Analysis}
\begin{figure}[t]
    \centering
    \includegraphics[width=\linewidth]{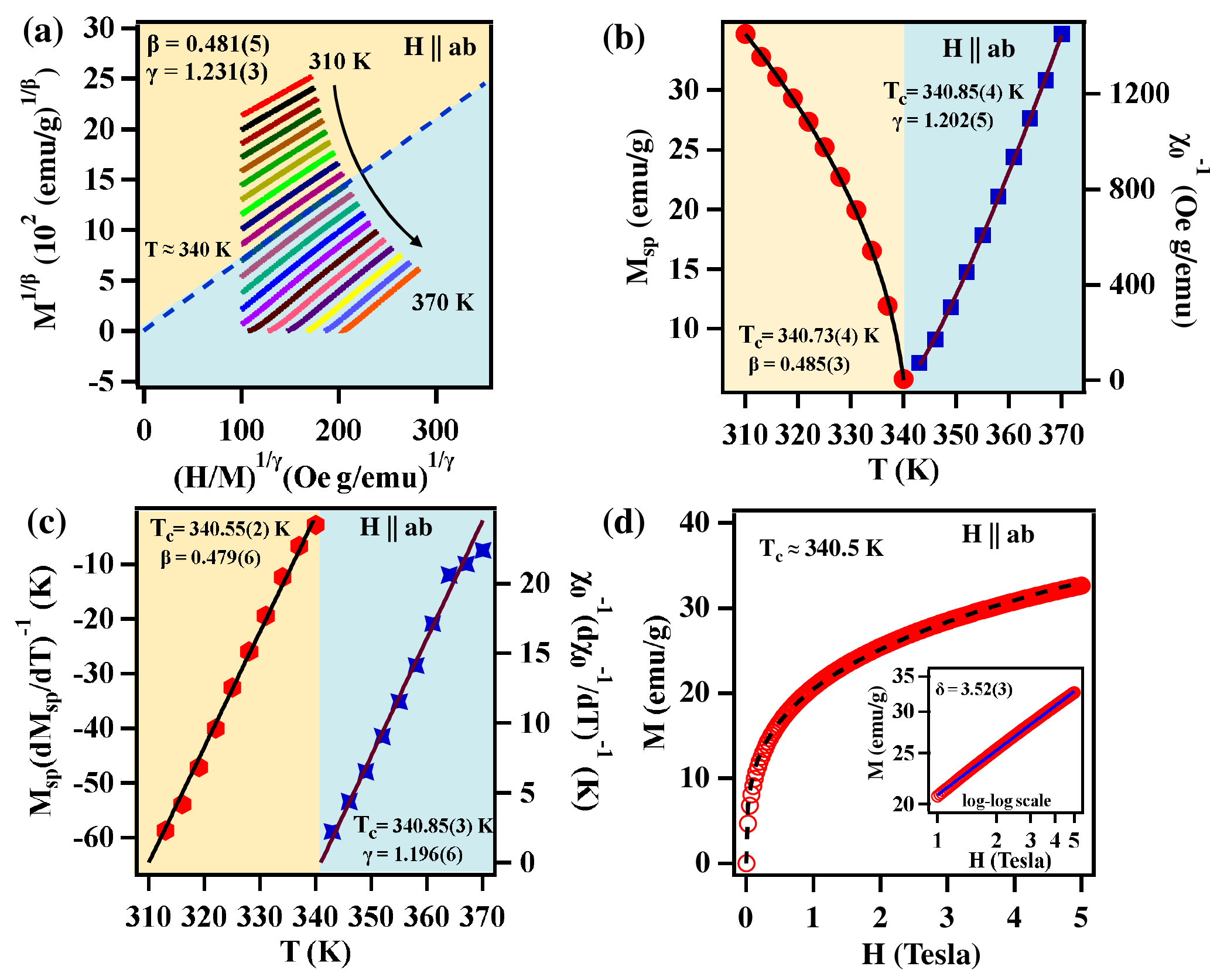}
     \caption{ (a) Modified Arrott plot of $M^{1/\beta}$ vs $(H/M)^{1/\gamma}$ at high field values. (b) Temperature dependent spontaneous magnetization $M_{sp}(T)$ (left-axis) and inverse initial susceptibility $\chi_0^{-1}(T)$ (right-axis). (c) Kouvel-Fisher plot of temperature dependent $M_{sp}(T)[dM_{sp}(T)/dT]^{-1}$ (left-axis) and $\chi_0^{-1}(T)[d \chi_0^{-1}(T)/dT]^{-1}$ (right-axis). (d) Isothermal magnetization $M(H)$ at $T_C$ $\approx$ 340.5 K. Inset in (d) shows $M(H)$ plot in the log-log scale.}
    \label{5}
\end{figure}

Scaling hypothesis suggests that the second order phase transition in a magnetic system around T$_C$ can be described by a set of critical exponents and magnetic equations of the state~\cite{stanley1971phase}.  Thus, the spontaneous magnetization $M_{sp}$ below $T_C$, inverse susceptibility $\chi_0^{-1}$ above $T_C$, and magnetization isotherms $M(H)$ at $T_C$ are characterized by the critical exponents $\beta$, $\gamma$, and $\delta$, respectively.  Mathematical equations governing the critical exponents are given by,

\begin{equation}
M_{sp}(T) = M_0(-\epsilon)^\beta \text{ for } \epsilon < 0, T < T_c,
\label{Eq1}
\end{equation}

\begin{equation}
\chi_0^{-1}(T) = (h_0/m_0)\epsilon^\gamma, \text{ for } \epsilon> 0, T > T_c
\label{Eq2}
\end{equation}

\begin{equation}
 M = DH^{1/\delta}, \text{ for } \epsilon = 0, T = T_c,
 \label{Eq3}
\end{equation}

where $\epsilon$ = ($T$ – T$_C$)/T$_C$ is the reduced temperature, $M_0$, $h_0/m_0$, and $D$ are the critical amplitudes\cite{fisher1967theory}. In order to obtain the critical exponents $\beta$, $\gamma$, and $\delta$ as well as the exact value of T$_C$,  we first did the Arrott-plot $M^{2}$ vs. $H/M$ as shown in Fig.~\ref{4}(a). However, this method does not produce the required parallel linear curves at higher fields, particularly below $T_C$~\cite{PhysRev.108.1394}. Therefore, we used the modified Arrott-plot (MAP) technique and plotted  $M^{1/\beta}$ vs. $(H/M)^{1/\gamma}$  for various models around T$_C$. Fig~.\ref{4}(b) shows MAP for 3D XY model ($\beta = 0.345, \gamma = 1.316$)~\cite{PhysRevB.21.3976}, Fig.~\ref{4}(c) shows MAP for 3D Ising model ($\beta = 0.325, \gamma = 1.241$)~\cite{kaul1985static}, Fig.~\ref{4}(d) shows MAP for 2D Ising model ($\beta = 0.125, \gamma = 1.75$)~\cite{PhysRevLett.29.917}, Fig.~\ref{4}(e) shows MAP for  3D Heisenberg model ($\beta = 0.365, \gamma = 1.386$)~\cite{kaul1985static}, and Fig.~\ref{4}(f) shows  MAP for Tricritical meanfield model ($\beta = 0.25, \gamma = 1$)~\cite{PhysRevLett.89.227202}. As can be seen from Figs.~\ref{4}(b)-\ref{4}(f), none of these models can actually produce the required parallel linear curves at higher magnetic fields from their respective critical exponents, confirming that a single theoretical model cannot properly describe the magnetic interactions present in Cr$_4$Te$_5$. Thus, we used the modified Arrott-plot technique in an iterative method in the vicinity of $T_C$ to produce parallel linear curves as shown in Fig.~\ref{5}(a)~\cite{PhysRevLett.19.786,PhysRevB.79.214426}.   This procedure gives $\chi_0^{-1}(T)$ and $M_{sp}(T)$ as the intercepts on the $x$-axis and $y$-axis, respectively. Fig.~\ref{5}(b) depicts $M_{sp}$ and $\chi_0^{-1}$ plotted as a function of temperature derived from Fig.~\ref{5}(a). From Eqs.~\ref{Eq1} and \ref{Eq2} we estimate the critical exponents $\beta$ = 0.485(3) with $T_C$ = 340.73(4) K and $\gamma$ = 1.202(5) with $T_C$ = 340.85(4) K. The estimated $T_C$ values are very close to the value of 338 K obtained from the $M(T)$ data [see Figs.~\ref{2}(a-b)].  Further, the critical exponents can be found more accurately using the Kouvel-Fisher (KF) plots~\cite{PhysRev.136.A1626}, governed by the below equations.

\begin{equation}
M_{sp}(T)[dM_{sp}(T)/dT]^{-1} = (T - T_c)/\beta,
\label{Eq4}
\end{equation}

\begin{equation}
 \chi_0^{-1}(T)[d \chi_0^{-1}(T)/dT]^{-1} = (T - T_c)/\gamma,
 \label{Eq5}
\end{equation}

Fig.~\ref{5}(c) depicts the Kouvel-Fisher plots fitted with the Eqs.~\ref{Eq4} and \ref{Eq5} below and above $T_C$. From the KF method fitting we derive the critical exponents $\beta$ = 0.479(6) with $T_C$ = 340.55(2) K and $\gamma$ = 1.196(6) with $T_C$ = 340.85(3) K. The values of $\beta$ and $\gamma$ obtained from the KF method are in good agrement with the values obtained from the MAP method. In addition, the exponent $\delta$ has been calculated using the Widom scaling relation, $\delta$ = 1 + $(\gamma$/$\beta)$~\cite{widom1964degree}. Thus, $\delta$ = 3.478 and $\delta$ = 3.497 are calculated from MAP and KF plots, respectively. Importantly, the $\delta$ values obtained from both MAP and KF methods are close to the value $\delta$ = 3.52(3) obtained using the critical isotherm (CI) following the Eq.~\ref{Eq3} as shown in Fig.~\ref{5}(d). All the experimentally derived critical exponents are summarized in the Table~\ref{T2}. Note here that, although the critical exponents obtained from MAP and KF techniques are close, we used the critical exponents of MAP for further discussions.  Therefore, the critical exponent $\beta$=0.485(3) obtained from this study is close to the meanfield-model value of $\beta$=0.5 and $\gamma$=1.202(5) is close to the 3D Heisenberg-model value of $\gamma$=1.386.

%  while the critical exponent $\gamma$=1.196(6) is close to the 3D Heisenberg-model value of $\gamma$=1.2. Such a mixed magnetic interactions are possibly due to the strong out-of-plane (Cr-Cr) magnetic interactions along the $c$-axis~\cite{Dijkstra_1989}. Also, a recent another study on Cr$_4$Te$_5$ came to a similar conclusion that the system has mixed magnetic interactions of meanfield-like and 3D Heisenberg-like~\cite{Zhang2020}.

\begin{figure}[t]
    \centering
    \includegraphics[width=\linewidth]{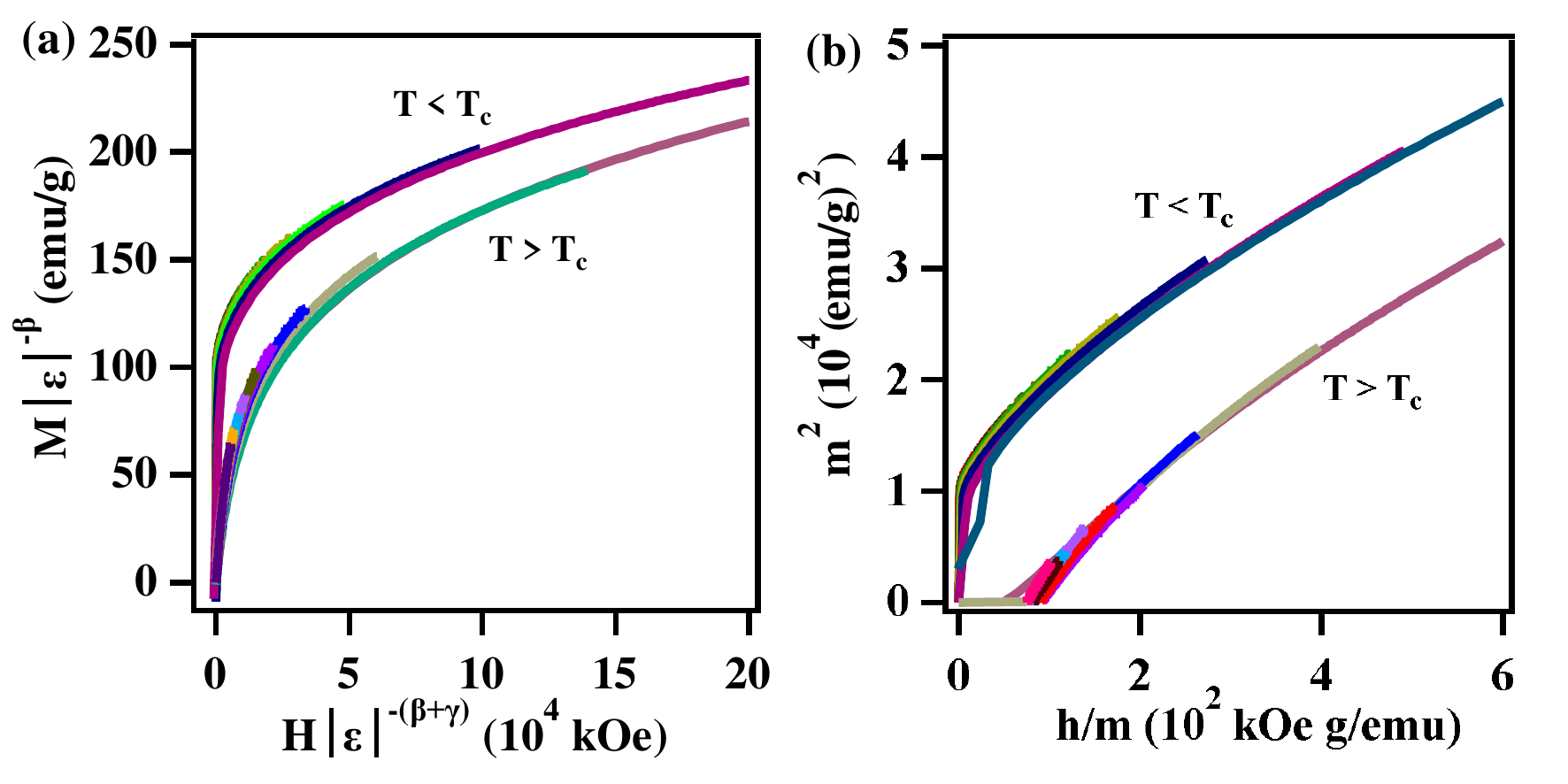}
    \caption{ (a) and (b) are the normalized magnetization M and m$^2$ plotted as a function of re-normalized field $h$ below and above T$_C$. See the text for more details.}
    \label{6}
\end{figure}

Further, to verify the accuracy of critical exponents and critical temperature obtained from MAP and KF methods, we employed scaling analysis~\cite{stanley1971phase}. The scaling equations of the state for magnetic systems are given by,
 \begin{equation}
        M (H, \epsilon) = \epsilon^\beta f_\pm (H/\epsilon^{\beta + \gamma})
        \label{Eq6}
 \end{equation}
 Above equation can be write as well,
\begin{equation}
    m = f_\pm (h)
     \label{Eq7}
\end{equation}

where $m = \epsilon^{-\beta} M (H, \epsilon) $ is the normalized magnetization and $h =  H\epsilon^{-(\beta + \gamma)}$ is the normalized field.
For the correct $\beta$ and $\gamma$ values obtained from the MAP and KF methods, following the scaling hypothesis, the scaled magnetization $m$ and field $h$ should fall on two different branches of the universal curves when plotted using the Eq.~\ref{Eq6} and Eq.~\ref{Eq7}. One branch for T$<$T$_C$ and another branch for T$>$T$_C$, as shown in Figs.~\ref{6}(a) and \ref{6}(b). Since all the curves shown in Figs.~\ref{6}(a) and \ref{6}(b) overlap with respective universal curves, the scaling hypothesis reaffirms the reliability of the derived critical exponents using the MAP and KF methods.

 \begin{table}[h]
\caption{Obtained critical parameters following the renormalization group theory analysis}
\begin{tabular*}{\linewidth}{c @{\extracolsep{\fill}} ccccc}
 \hline
\{d : n\} & $\gamma$ & $\beta$ & $\sigma$ & $\delta$ \\ [1.5ex]
 \hline\hline
\{2 : 1\} & 1.20 & 0.38 & 1.22 & 4.158  \\ [1.2ex]
 \hline
\{2 : 2\} & 1.20 & 0.41 & 1.19 & 3.927 \\[1.2ex]
 \hline
\{2 : 3\} & 1.20 & 0.55 & 1.05 & 3.182\\[1.2ex]
 \hline
\{3 : 2\} & 1.20 & 0.405 & 1.79 & 3.963\\[1.2ex]
 \hline
 \{3 : 3\} & 1.20 & 0.46 & 1.71 & 3.609 \\[1.2ex]
 \hline
\end{tabular*}
\label{T1}
\end{table}

Next, the renormalization group (RG) theory analysis suggest that the long- and short- range exchange interactions decay with the distance $r$ as $J (r) = r^{-(d + \sigma)}$ and $J (r) = e^{-r/b}$, respectively~\cite{PhysRevLett.29.917}. Here, $d$ is the spatial dimensionality, $\sigma$ is a positive constant, and $b$ is spatial scaling factor.  The relation between the critical exponent $\gamma$ and $\sigma$ is given by,

 \begin{equation}
 \begin{split}
  \gamma & = 1 + \frac{4}{d}\frac{(n + 2)}{(n + 8)}\Delta\sigma +\frac{8(n - 4)(n + 2)}{(n + 8)^2d^2}\\
         &  * [\frac{2(7n + 20)G(d/2)}{(n + 8)(n - 4)}+1 ]\Delta\sigma^2
     \end{split}
     \label{Eq8}
 \end{equation}

 where $\Delta\sigma = \sigma - d/2$ , $ G(d/2) = 3 -0.25*(d/2)^2$. $n$ and $d$ are the spin and spatial dimensionality of the system. By considering the experimental $\gamma$ value of 1.196(6), following the RG theory, we have calculated the critical exponent $\beta$ and local exponent $\sigma$ for various sets of \{d : n\} values and are tabulated in Table~\ref{T1}.  From this, we find that the critical exponent $\beta$ and $\delta$ obtained using the 3D Heisenberg-model of  \{d = 3 : n = 3\} are close to the experimentally obtained values of the critical exponents.  Thus, the long-range exchange interactions in Cr$_4$Te$_5$ decays as $J (r) = r^{-(d + \sigma)}= r^{-4.71}$ for $d=3$ and $\sigma$=1.71.

\subsection{Magnetocaloric Effect}

In order to explore the magnetocaloric effect (MCE), we analyzed the field dependent isotherms $M(H)$ taken at different temperatures [see Figs.~\ref{3}(a) and \ref{3}(b)]  in the vicinity of $T_C$ for both  H$\parallel ab$ and H$\parallel c$ orientations. Magnetocaloric effect is an intrinsic property of a ferromagnetic system, causing heating or cooling adiabatically under the applied magnetic fields~\cite{PECHARSKY199944}. Thus, a magnetic entropy change $\Delta S_m (T, H)$ is induced in the presence of magnetic fields which is represented by the formula,
\begin{equation}
\Delta S_m (T, H) =  \int_{o}^{H} (\frac{\partial S}{\partial H})_T dH = \int_{o}^{H} (\frac{\partial M}{\partial T})_H dH
 \label{Eq9}
 \end{equation}
where ($\frac{\partial S}{\partial H})_T$ = ($\frac{\partial M}{\partial T})_H$  based on Maxwell’s relation. In case of magnetization measured at small discrete field and temperature intervals, $\Delta S_m (T, H)$ could be written as
\begin{equation}
\Delta S_m (T, H) = \frac{\int_{0}^{H} M(T_{i+1}, H) dH - \int_{0}^{H} M(T_{i}, H) dH }{T_{i+1}-T_i}
 \label{Eq10}
\end{equation}

\begin{figure*} [t]
    \centering
    \includegraphics[width=0.95\linewidth]{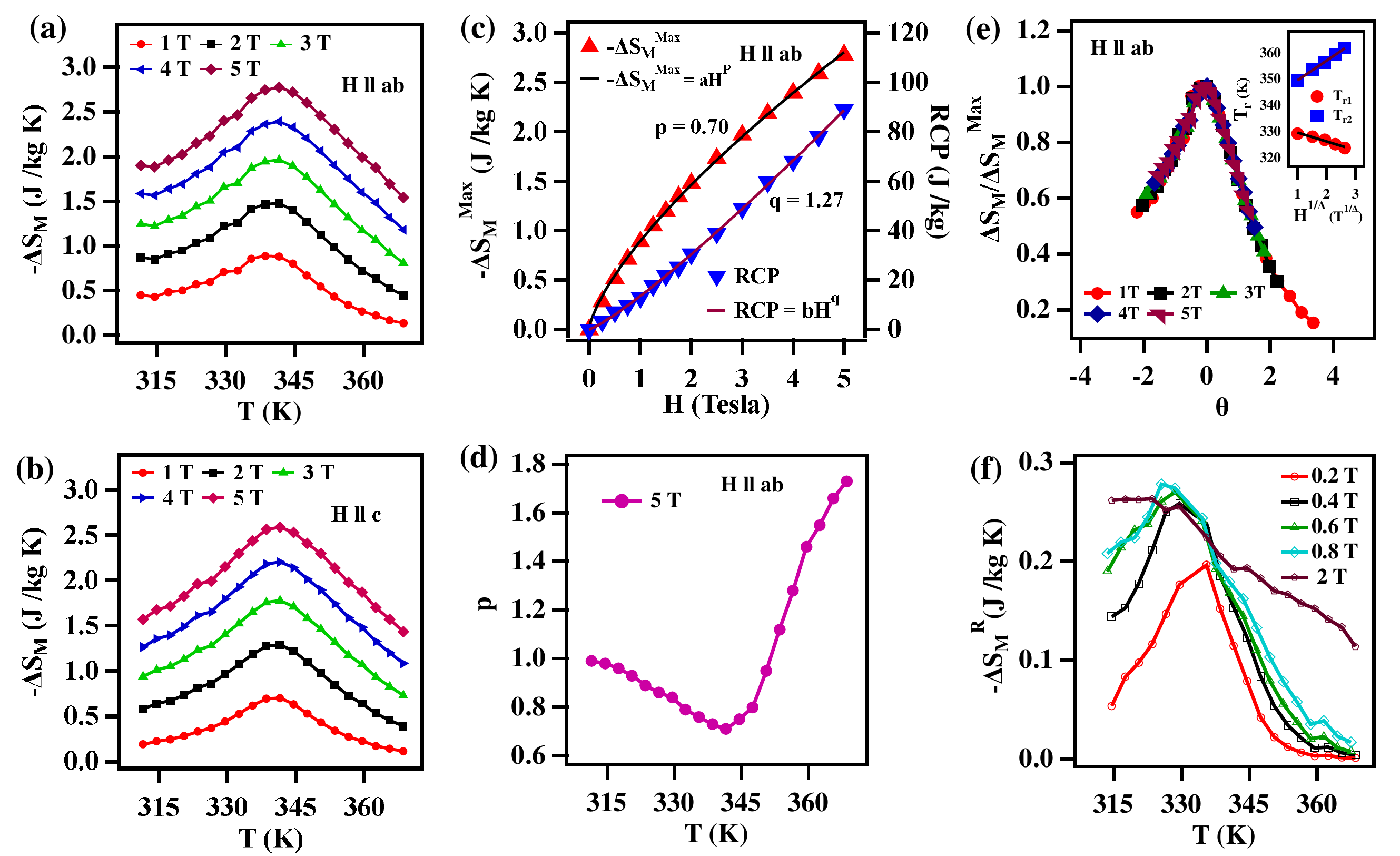}
    \caption{The calculated magnetic entropy -$\Delta S_m$ as a function of temperature at different magnetic field. (a) H $||$ ab, (b) H $||$ c, (c) Field dependence of maximum magnetic entropy changes (-$\Delta S_M^{max}$) (left axis) and relative cooling power (RCP) (right axis), (d) The exponent p as function of temperature obtained from the fitting of field dependent isothermal magnetic entropy change at various temperature, (e) The normalised magnetic entropy changes as a function of rescaled temperature $\theta$ at different field, Inset shows T$_r$ vs H$^{1/\Delta}$ with $\Delta$ = $\beta$ + $\gamma$, (f) Rotational magnetic entropy change (-$\Delta S_M^R$) as a function of temperature.}
    \label{7}
\end{figure*}

Figs.~\ref{7}(a) and \ref{7}(b) show - $\Delta S_m (T, H)$ plotted as a function temperature under various magnetic fields up to 5 T taken with a step size of 1 T for both H$\parallel ab$ and H$\parallel c$ orientations, respectively. All - $\Delta S_m (T, H)$ curves show a maximum change in entropy with a broad peak at around $T_C$ as seen from Figs.~\ref{7}(a) and \ref{7}(b). Further, we observe that the value of - $\Delta S_m (T, H)$ and RCP increase monotonically with field for $H\parallel ab$ as shown in Fig.~\ref{7}(c). Under an applied field of 5 T, the maximum of -$\Delta S_m (T, H)$ is about 2.78 J kg$^{-1}$ K$^{-1}$ for H$\parallel ab$ and is about 2.58 J kg$^{-1}$ K$^{-1}$  for H$\parallel c$. These -$\Delta S_m (T, H)$ values taken at 5 T are comparable to the other 2D ferromagnetic systems such as Cr$_2$Ge$_2$Te$_6$ (2.64 J kg$^{-1}$ K$^{-1}$)~\cite{PhysRevMaterials.3.014001} and Cr$_5$Te$_8$ (2.38 J kg$^{-1}$ K$^{-1}$)~\cite{PhysRevB.100.245114}, larger than the values of Fe$_{3-x}$GeTe$_2$ (1.14 J kg$^{-1}$ K$^{-1}$)~\cite{Liu2019} and CrI$_3$ (1.56J kg$^{-1}$ K$^{-1}$)~\cite{Liu2018}, and smaller than the values of CrB$_3$ (7.2 J kg$^{-1}$ K$^{-1}$)~\cite{Yu2019} and Cr$_2$Si$_2$Te$_6$ (5.05 J kg$^{-1}$ K$^{-1}$)~\cite{PhysRevMaterials.3.014001}. Further, -$\Delta S_m (T, H)$ of the studied Cr$_4$Te$_5$ are in good agreement with the earlier reported values of 2.58 J kg$^{-1}$ K$^{-1}$ and 2.42 J kg$^{-1}$ K$^{-1}$ for  H$\parallel ab$ and  H$\parallel c$, respectively~\cite{Zhang2020}.

%Fe$_3$GeTe$_2$ ()\cite{doi:10.1021/acs.inorgchem.5b01260},

To estimate the relative cooling power (RCP), we employed the relation RCP = -$\Delta S_M^{max}$ $\times$ $\delta T_{FWHM}$, where -$\Delta S_M^{max}$ is the maximum entropy change near $T_C$ and $\delta T_{FWHM}$ is the full width at half maximum of the peak~\cite{gschneidner1999recent}. The calculated RCP in Cr$_4$Te$_5$ is of 88.29 J kg$^{-1}$ at around $T_C$ with an applied field of 5 T parallel to $\textit{ab}$-plane. The RCP value of Cr$_4$Te$_5$  obtained in this study  is comparable to the RCP value obtained in the other 2D systems such as  Cr$_2$Ge$_2$Te$_6$ (87 J kg$^{-1}$)\cite{PhysRevMaterials.3.014001}, but smaller than the values obtained from Cr$_5$Te$_8$ (131.2 J kg$^{-1}$)~\cite{PhysRevB.100.245114}, CrI$_3$ (122.6 J kg$^{-1}$)~\cite{Liu2018}, Cr$_2$Si$_2$Te$_6$ (114 J kg$^{-1}$)~\cite{PhysRevMaterials.3.014001}, Fe$_{3-x}$GeTe$_2$(113 J kg$^{-1}$)~\cite{Liu2019}, and CrBr$_3$(191.5 J kg$^{-1}$)~\cite{Yu2019}.

In addition,  both -$\Delta S_M^{max}$ and RCP are related by the power law of magnetic field as given below~\cite{gschneidner1999recent,franco2006field},
\begin{equation}
  -\Delta S_M^{max} = aH^p
   \label{Eq11}
 \end{equation}
 \begin{equation}
      RCP =  bH^q
       \label{Eq12}
 \end{equation}
 where p and q are the exponents. At T=T$_C$,  they can be written as
 \begin{equation}
     p = 1 + \frac{\beta - 1}{\beta + \gamma}
      \label{Eq13}
 \end{equation}
 \begin{equation}
     q = 1 + \frac{1}{\delta}
      \label{Eq14}
 \end{equation}

  \begin{table}[h]
\caption{Critical exponents of Cr$_4$Te$_5$. The MAP, KF, and CI represent modified Arrott plot, the Kouvel-Fisher plot, and critical isotherm, respectively. }
\begin{tabular*}{\linewidth}{c @{\extracolsep{\fill}} ccccc}
 \hline
 Technique & $\beta$ & $\gamma$ & $\delta$ & p & q \\ [1.5ex]
 \hline\hline
-$\Delta S_M^{max}$ &  &  &  & 0.70 & \\ [1.2ex]
 \hline
RCP &  &  & & & 1.27 \\[1.2ex]
 \hline
MAP & 0.485 & 1.202 & 3.472 & 0.695 & 1.288 \\[1.2ex]
 \hline
KFP & 0.479 & 1.196 & 3.515 & 0.683 & 1.285 \\[1.2ex]
 \hline
 CI &  &  & 3.52 & &1.284 \\ [1.2ex]
 \hline
\end{tabular*}
\label{T2}
\end{table}

Fig.~\ref{7}(c) summarizes the field dependence of -$\Delta S_M^{max}$ and RCP.  The fit of -$\Delta S_M^{max}$ by the Eq.~\ref{Eq11} yields $p$=0.70. This is in very good agreement with the $p$ value of 0.689(4) obtained from the Eq.~\ref{Eq13}.  Similarly, the field dependence of  RCP is fitted by the Eq.~\ref{Eq12}, yielding to  $q$ = 1.27 that is close to the value of 1.28(1) obtained from the Eq.~\ref{Eq14} using the critical exponent of $\delta$  derived from the magnetic isotherms. Fig.~\ref{7}(d) displays the temperature dependence of $p$ at 5 T. The $p(T)$ curve follows universal behaviour across $T_C$ as it reaches to the value 1 for T $<$ $T_C$~\cite{FRANCO20091115}. On other hand, well above $T_C$, $p$ reaches to 2 as a consequence of the Curie-Weiss law~\cite{FRANCO20091115}. At T = $T_C$, $p(T)$ has a minimum value of 0.7, consistent with the minimum value of 0.75 as per the universal curve~\cite{franco2006field}. Overall, temperature dependence of $p$ perfectly follows the universal behaviour of a second order phase transition~\cite{franco2006field}.

Next,  we performed the scaling analysis of MCE following the procedure given by Franco \textit{et. al.} ~\cite{franco2006field,franco2007constant}. The scaling analysis of - $\Delta S_m (T, H)$ is constructed by normalizing the - $\Delta S_m (T, H)$ curves with respect to the maximum of -$\Delta S_M^{max}$ [$\frac{\Delta S_m (T, H)}{\Delta S_M^{max}}$]. The reduced temperature ($\theta_{\mp}$) is defined by choosing two reference temperatures ($T_{r1} \le T_C$ and $T_{r2}> T_C$), satisfying the condition, $\frac{\Delta S_m (T_{r1} < T_c)}{\Delta S_M^{max}}$  = $\frac{\Delta S_m (T_{r2} > T_c)}{\Delta S_M^{max}}$ = h. Here, h is a scaling constant having the values within the range of $0 < h < 1$. Then, the rescaled temperature $\theta_{\mp}$ can be written as

 \begin{align}\label{Eq15}
   \theta_{-}  = (T_C - T)/(T_{r1} - T_C),T \le T_C\\
   \theta_{+} = (T - T_C)/(T_{r2} - T_C), T > T_C
  \end{align}

 Following MCE scaling analysis, all the curves of $\Delta S_m/\Delta S_M^{max}$ plotted as a function of reduced temperature $\theta$ at various magnetic fields collapse into a single curve as shown in Fig.\ref{7}(e). This, confirms the second order magnetic phase transition in Cr$_4$Te$_5$~\cite{oesterreicher1984magnetic,law2018quantitative,franco2017predicting}.  Moreover, it can be seen that the universal curve of MCE is independent of the applied magnetic field and temperature as it generally determined by the intrinsic magnetization of the system~\cite{franco2006field}. Next, the reference temperatures $T_{r1}$ and $T_{r2}$ linearly depends on $H^{1/\Delta}$ with $\Delta$ = $\beta$ + $\gamma$ as shown in the inset of Fig.~\ref{7}(e). The rotational magnetic entropy changes ($\Delta S_M^R$) can be calculated using the formula $\Delta S_M^R(T, H)$ = $\Delta S_m(T, H_{ab}) - \Delta S_m(T, H_{c})$. Fig.~\ref{7}(f) depicts the temperature dependence of - $\Delta S_M^R(T, H)$. From Fig.~\ref{7}(f) we can find a maximum of -$\Delta S_M^R(T, H)$ at around T$_C$ when derived at a field interval of 0.2 T. The maxima of -$\Delta S_M^R(T, H)$ shifts to lower temperatures with increasing field intervals and for 2 T of field interval no maxima is found down to 315 K. This behaviour is generally found in the systems with strong magnetic anisotropy~\cite{PhysRevB.100.245114}.

 Finally, before concluding,  we would like to compare our results with the existing literature on the composition of Cr$_4$Te$_5$. Till date there exits two studies on this system, (i) in the bulk phase~\cite{Zhang2020} and (ii) in thinfilm phase~\cite{Wang2022}. The Curie temperature of 338 K observed from our $M(T)$ study is higher than the Curie temperature of $\approx$ 320 K previously reported in both the bulk and thinfilm phases. It is known that in Cr$_x$Te$_y$ systems the Curie temperature depends on various parameters such as sample thickness,  sample composition, and on the method of preparation~\cite{Li2019, Lee2021}. The critical exponents of  $\beta$=0.485(3) and $\gamma$=1.202(5) are close to the meanfield-like and 3D Heisenberg-like interactions, respectively,  while the critical exponents of  $\beta$=0.388 and $\gamma$=1.29 as reported in Ref.~\cite{Zhang2020} in the bulk phase are found to be close to the 3D Heisenberg-like  and meanfield-like interactions, respectively. On the other hand, the critical exponent found in the thinfilm phase $\beta$=0.491 is close to the meanfield-like interactions, consistent with our observations~\cite{Wang2022}. Such a mixed phase interactions are reported in many other 2D Ferromagnets such as in Fe$_{0.26}$TaS$_2$ ($\beta$ = 0.459, $\gamma$ = 1.205)~\cite{Zhang2019} and Co$_{0.22}$TaS$_2$ ($\beta$ = 0.43, $\gamma$ = 1.15)~\cite{Liu2021}. Nevertheless, the renormalization group theory analysis suggests the 3D Heisenberg-type magnetic interactions in the studied system when estimated using the experimental $\gamma$ value of 1.202.

\section{Summary}
To summarize, we thoroughly studied the magnetic anisotropy and magnetocaloric effect associated with critical exponent analysis of Cr$_4$Te$_5$. The critical exponents of $\beta$, $\gamma$, and $\delta$ obtained from the modified Arrot-plot and Kouvel-Fisher techniques match very well. Scaling analysis of the magnetic entropy change confirms the second order magnetic phase transition in this system. Strong uniaxial magnetocrystalline anisotropy found in this system could stabilize the long-range FM ordering within few layers. The renormalization group theory analysis suggests 3D Heisenberg-type magnetic interactions in Cr$_4$Te$_5$.

\section{Acknowledgements}

Authors thank SERB (DST), India for financial support (Grant No. SRG/2020/000393). S.C. and S.G. acknowledge University Grants Commission (UGC), India for the PhD fellowship.

\bibliography{main}% Produces the bibliography via BibTeX.

\end{document}